\documentclass[fleqn,usenatbib]{mnras}

\usepackage{mathptmx}

\usepackage[T1]{fontenc}

\DeclareRobustCommand{\VAN}[3]{#2}
\let\VANthebibliography\thebibliography
\def\thebibliography{\DeclareRobustCommand{\VAN}[3]{##3}\VANthebibliography}

\usepackage{graphicx}	
\usepackage{amsmath}	
\usepackage{amssymb}	
\usepackage{natbib}	
\usepackage{threeparttable}
\usepackage{booktabs}
\usepackage{amsmath}
\usepackage{hyperref}

\title[$\alpha$-depleted BaSTI stellar models]{The updated BaSTI stellar evolution models and isochrones. IV. $\alpha$-depleted calculations}

\author[Pietrinferni et al.]{
Adriano Pietrinferni,$^{1}$\thanks{E-mail: adriano.pietrinferni@inaf.it}
Maurizio Salaris,$^{2,1}$
Santi Cassisi,$^{1,3}$
Alessandro Savino,$^{4}$
Alessio Mucciarelli,$^{5,6}$
\and
David Hyder,$^{2}$
Sebastian Hidalgo$^{7,8}$
\\
$^{1}$INAF -- Osservatorio Astronomico di Abruzzo, Via M. Maggini, s/n, I-64100, Teramo, Italy\\
$^{2}$Astrophysics Research Institute, Liverpool John Moores University, 146 Brownlow Hill, Liverpool, L3 5RF, UK\\
$^{3}$INFN -- Sezione di Pisa, Largo Pontecorvo 3, 56127 Pisa, Italy\\
$^4$ Department of Astronomy, University of California, Berkeley, Berkeley, CA, 94720, USA\\
$^5$ Dipartimento di Fisica e Astronomia “Augusto Righi”, Alma Mater Studiorum, Università di Bologna, Via Gobetti 93/2, I-40129 Bologna, Italy\\
$^{6}$INAF -- Osservatorio di Astrofisica e Scienza dello Spazio di Bologna, Via Gobetti 93/3, I-40129, Bologna, Italy\\
$^{7}$Instituto de Astrofisica de Canarias, Via Lactea s/n, La Laguna, Tenerife, Spain\\
$^{8}$Department of Astrophysics, University of La Laguna, Via Lactea s/n, La Laguna, Tenerife, Spain
}

\date{Accepted XXX. Received YYY; in original form ZZZ}

\pubyear{2015}

\begin{document}
\label{firstpage}
\pagerange{\pageref{firstpage}--\pageref{lastpage}}
\maketitle

\begin{abstract}
This is the fourth paper of our new release of the 
BaSTI (a Bag of Stellar Tracks and Isochrones) stellar model and isochrone library. Following the updated solar-scaled, $\alpha$-enhanced, and 
white dwarf model libraries, we present here $\alpha$-depleted ([$\alpha$/Fe] = $-$0.2) evolutionary tracks and isochrones, suitable to study the $\alpha$-depleted stars discovered in Local Group dwarf galaxies and in the 
Milky Way. These calculations include all improvements and updates 
of the solar-scaled and $\alpha$-enhanced models, and span a mass range between 0.1 and 15 $M_{\odot}$, 21 metallicities between [Fe/H] = -3.20 and +0.45 with a helium-to-metal enrichment ratio $\Delta Y$/$\Delta Z$ = 1.31, homogeneous with the solar-scaled and $\alpha$-enhanced models. 
The isochrones --available in several photometric filters-- cover an age range between $\sim$ 20 Myr and 14.5 Gyr, including the pre-main-sequence phase. We have compared our isochrones with independent calculations of $\alpha$-depleted stellar models, available for the same $\alpha$-element depletion adopted in present investigation. We have also discussed the effect of an $\alpha$-depleted heavy element distribution on 
the bolometric corrections in different wavelength regimes.
Our $\alpha$-depleted evolutionary tracks and isochrones are publicly available at the BaSTI website.
 \end{abstract}




\section{Introduction}

Extended (in terms of mass and chemical composition ranges) and accurate sets of stellar model calculations are fundamental inputs to a wide variety of numerical tools employed to interpret spectroscopic and photometric observations of individual stars, star clusters, and galaxies, both resolved and unresolved.

In the last years we have undertaken a massive update --in terms of physics inputs and parameter space coverage-- of our 
BaSTI (a Bag of Stellar Tracks and Isochrones) stellar models and isochrones  library, and published in Paper~I \citep{bastiiacss} a new grid of solar-scaled 
models and isochrones, in Paper~II its $\alpha$-enhanced counterpart \citep{bastiiacae}, followed in Paper~III by the extension of the library to the white dwarf regime \citep{bastiwdnew}\footnote{All these libraries are publicly available at the new 
official BaSTI website \url{http://basti-iac.oa-abruzzo.inaf.it}.}.

In this paper (Paper~IV of this series) we present a further extension of the new BaSTI library to cover chemical compositions with an $\alpha$-depleted ([$\alpha$/Fe]=$-$0.20) metal distribution. To the best of our knowledge, there are only two other stellar model and isochrone libraries for $\alpha$-depleted chemical mixtures in the literature, published by \citet{dartmouth} and \citet{vand14}. 

The presence of $\alpha$-depleted stars has been revealed by spectroscopic observations of Local Group dwarf galaxies, as discussed in the classic review by \citet{tolstoy} -- see also the results by \citet{vargas} for a sample of ultra-faint dwarfs, and \citet{hill} for the  Sculptor dwarf galaxy -- and in the 
Milky Way \citep{nissen, hayes, lamost}. An accurate characterization of these populations of stars can indeed greatly benefit from stellar models and isochrones with the appropriate metal abundance distribution, given the sensitivity of the bolometric corrections at optical and shorter wavelength to the specific metal abundance pattern \citep[see][and the next sections]{bcaen}.

This paper is organized as follows. Section~\ref{inp} summarizes the physics inputs and the heavy element distribution adopted in the calculations, 
followed by Sect.~\ref{model} which first  
presents the stellar model grid and its mass, chemical composition, evolutionary phases coverage, and then shows   
comparisons with previous calculations available in the literature. 
Section~\ref{bolom} discusses the importance of using 
bolometric corrections calculated for the appropriate  
$\alpha$-depleted metal mixture when comparing models and isochrones to 
data in colour-magnitude diagrams, and a summary follow in Section~\ref{concl}.

\section{Metal distribution and physics inputs}
\label{inp}

For these calculations, we have employed the same stellar evolution code and physics inputs of Paper~I and II.   
The adopted $\alpha$-depleted metal distribution listed in Table~\ref{tab:mix} has been calculated with the $\alpha$-elements O, Ne, Mg, Si, S, Ca, and Ti uniformly depleted with respect to Fe in comparison to the corresponding solar ratios \citep[the reference solar-scaled metal distribution is from][as in Paper~I and II]{caffau} by 0.2~dex ([$\alpha$/Fe]=$-$0.2).

This metal distribution has been employed consistently for the calculation of the nuclear energy generation, radiative and electron conduction opacities, as well as in the equation of state (and bolometric corrections, see Sect.~\ref{bolom}). The sources for all these physics inputs are given in Paper I, together with details about the treatment of the neutrino energy loss rates, superadiabatic convection (we employ a mixing length $\alpha_{\rm ML}$ = 2.006 obtained from a solar model calibration), overshooting from the convective cores, and atomic diffusion (without radiative levitation). As for the outer boundary conditions, in the case of models with initial mass larger than about $0.5~M_\odot$ we used 
the integration of a $T(\tau)$ relation as in Paper~I and II. At lower masses, i.e. in the range of very low-mass stars, for both solar-scaled and $\alpha$-enhanced models we used outer boundary conditions provided by model atmosphere computations (see Paper~I and II for details). 
However, for our $\alpha$-depleted calculations we have not been able to find 
publicly available outer boundary conditions from model atmospheres with the appropriate
$\alpha$-depleted metal mixture, consistent with those adopted for the other sets of models in the BaSTI-IAC library. We were therefore forced to use also for 
very low-mass stars the integration of the same $T(\tau)$ relationship adopted for larger masses.

Mass loss has been included by means of the \citet{reimers} formula, 
with the free parameter $\eta$ set to 0.3 (see Paper~I and II for details).

\begin{table}
\centering
\caption{\label{tab:mix} The adopted $\alpha$-depleted heavy element mixture.}
{
\begin{tabular}{lcc}
\hline
Element & mass fraction & number fraction \\
\hline
C  & 0.238566 &	 0.34146\\
N  & 0.063738 &	 0.07823\\     
O  & 0.364857 &	 0.39204\\       
Ne & 0.093950 &	 0.08004\\      
Na & 0.002955 &	 0.00221\\       
Mg & 0.034963 &	 0.02473\\       
Al & 0.005117 &	 0.00326\\     
Si & 0.039486 &	 0.02417\\       
P  & 0.000559 &	 0.00031\\       
S  & 0.017940 &	 0.00962\\       
Cl & 0.000412 &	 0.00020\\       
Ar & 0.008319 &	 0.00358\\       
K  & 0.000318 &	 0.00014\\       
Ca & 0.003404 &	 0.00146\\       
Ti & 0.000167 &  0.00006\\       
Cr & 0.001543 &  0.00051\\       
Mn & 0.001118 &  0.00035\\       
Fe & 0.116168 &  0.03576\\       
Ni & 0.006419 &  0.00188\\      
\hline
\end{tabular}
}
\end{table}

\section{The model library}
\label{model}

To allow easy comparisons and interpolations with our solar-scaled and the $\alpha$-enhanced libraries, the $\alpha$-depleted models have been calculated 
for the same [Fe/H] values of the other two metal mixtures. For each assumed value of [Fe/H] we have derived the corresponding metallicity Z, and the associated He 
mass fraction (Y) adopting $dY/dZ$ the helium-enrichment ratio equal to 1.31
(see Paper~I). Our model grid includes 21 metallicity values, listed in Table~\ref{tab:z}.

\begin{table}
\centering
\caption{\label{tab:z} Grid of initial chemical abundances.}
{
\begin{tabular}{rrcc}
\hline
[Fe/H] & [M/H] & Z & Y \\
\hline
        0.45   &     0.33  &    0.03037  &    0.2867\\
        0.30   &     0.18  &    0.02210  &    0.2759\\
        0.15   &     0.03  &    0.01596  &    0.2679\\
        0.06   &    $-$0.06  &    0.01309  &    0.2641\\
       $-$0.08   &    $-$0.20  &    0.00959  &    0.2595\\
       $-$0.20   &    $-$0.32  &    0.00733  &    0.2566\\
       $-$0.30   &    $-$0.42  &    0.00585  &    0.2546\\
       $-$0.40   &    $-$0.52  &    0.00466  &    0.2531\\
       $-$0.60   &    $-$0.72  &    0.00296  &    0.2509\\
       $-$0.70   &    $-$0.82  &    0.00235  &    0.2501\\
       $-$0.90   &    $-$1.02  &    0.00149  &    0.2489\\
       $-$1.05   &    $-$1.17  &    0.00106  &    0.2484\\
       $-$1.20   &    $-$1.32  &    0.00075  &    0.2480\\
       $-$1.30   &    $-$1.42  &    0.00059  &    0.2478\\
       $-$1.40   &    $-$1.52  &    0.00047  &    0.2476\\
       $-$1.55   &    $-$1.67  &    0.00033  &    0.2474\\
       $-$1.70   &    $-$1.82  &    0.00024  &    0.2473\\
       $-$1.90   &    $-$2.02  &    0.00015  &    0.2472\\
       $-$2.20   &    $-$2.32  &    0.00007  &    0.2471\\
       $-$2.50   &    $-$2.62  &    0.00004  &    0.2471\\
       $-$3.20   &    $-$3.32  &    0.00001  &    0.2470\\
 \hline
\end{tabular}
}
\end{table}

For each initial chemical composition, evolutionary sequences for 56 different values of the initial stellar mass have been computed: the minimum initial mass is $0.1M_\odot$, while the maximum value is 15$M_\odot$. For initial masses below $0.2M_\odot$, we computed evolutionary tracks for masses equal to 0.10, 0.12, 0.15, and $0.18M_\odot$. In the range between 0.2 and $0.7M_\odot$ a mass step equal to 0.05$M_\odot$ has been adopted. Mass steps equal to $0.1M_\odot$, $0.2M_\odot$, $0.5M_\odot$, and $1M_\odot$ have been adopted for the mass ranges $0.7-2.6M_\odot$, $2.6-3M_\odot$, and $3-10M_\odot$, and masses larger than $10M_\odot$, respectively.

Models less massive than $4M_\odot$ have been computed from the pre-MS, whereas more massive models have been computed starting from the zero age main sequence configuration.

All stellar models --but those corresponding to very low-mass stars whose core H-burning lifetime is much longer than the Hubble time-- have been calculated until the start of the thermal pulses on the asymptotic giant branch, or C-ignition for the more massive ones. For the very low-mass stars, the calculations have been stopped when the central H mass fraction is $\approx0.3$ (corresponding to ages already much older than the Hubble time).
For each metallicity, an extended set of low-mass core He-burning models suitable for the study of the horizontal branch (HB) in old stellar populations has been also computed, as done in Paper~I and II.

All the evolutionary sequences have been normalized, i.e. reduced to the same number of lines using the technique of \lq{equivalent points\rq} as described in Paper~I and II.\footnote{Although the normalization procedure is exactly the same adopted in \citet{bastiiacss}, we have slightly modified the definition of one of the selected key points to better account for the
extremely long evolutionary timescales of very low-mass stars. More in detail, the fourth key point for these models now corresponds to the evolutionary stage when H-burning has decreased the central hydrogen mass fraction by 0.01 with respect to the initial value.} For consistency, the same criterion has been applied to the solar-scaled and $\alpha$-enhanced models in the library and used for computing isochrones in a wide age range.

Having evolutionary tracks and isochrones computed this way, with the same number of points together with the homogeneous [Fe/H] grid for solar-scaled, $\alpha$-enhanced and $\alpha$-depleted calculations, makes it easy to calculate evolutionary tracks and/or isochrones at fixed [Fe/H] for any arbitrary value of [$\alpha$/Fe] 
between -0.2 and +0.4~dex. One could use for example a parabola to interpolate in 
[$\alpha$/Fe] point-by-point along the tracks and isochrones.
The same is true also for interpolations in [Fe/H] at a given [$\alpha$/Fe].

Both evolutionary tracks and isochrones are provided in the theoretical HR diagram (HRD) and in various photometric systems as described in Paper~I and II,  by applying bolometric corrections obtained from theoretical spectra (consistent with those employed in Paper~I and II) calculated with the appropriate $\alpha$-depleted metal distribution. 
However, due to the lack of suitable $\alpha$-depleted synthetic spectra for VLM stars, in this mass regime we have been forced to adopt a different approach. For each value of [Fe/H] we have employed the solar-scaled bolometric corrections of Paper~I after applying - for each photometric filter - a shift in order to match the corresponding bolometric correction at $T_{\rm eff}\approx 5000~K$ and $\log{g}=4.5$ provided by the $\alpha$-depleted synthetic spectra obtained with the ATLAS9 code (see the discussion in Paper~I).

Several photometric systems have been added (or updated) since the publication of Paper~I, and Table~3 shows an updated list of the systems available in our model library for $\alpha$-enhanced, solar-scaled and, $\alpha$-depleted calculations.

\begin{figure}
            \includegraphics[width=\linewidth]{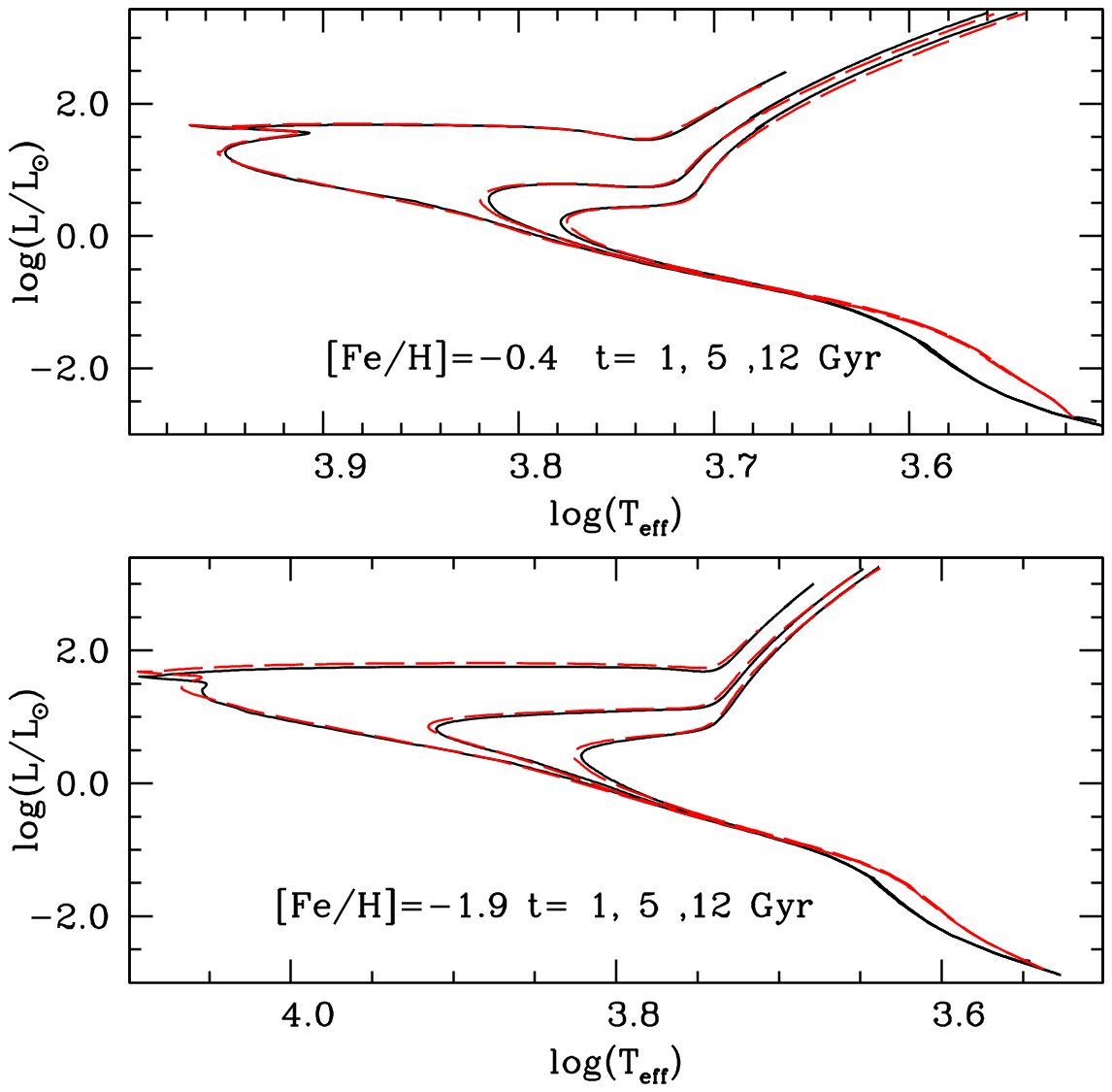}
    \caption{Comparison between \citet{dartmouth} and our 
    [$\alpha$/Fe]=$-$0.2 isochrones from the MS to the tip of the RGB (red dashed and black solid lines, respectively) with the labelled [Fe/H] and ages.}
    \label{fig_a}    
\end{figure}    


Figure~\ref{fig_a} shows the HRD of our 1, 5 and 12~Gyr isochrones for two values of [Fe/H], compared to the corresponding 
\citet{dartmouth} $\alpha$-depleted isochrones \citep[we could not compare with the calculations by][because they are for a 0.4~dex $\alpha$ depletion]{vand14}, from the main sequence (MS) to the tip of the red giant branch (RGB)\footnote{These isochrones have been computed using the online interpolation tool at \url{http://stellar.dartmouth.edu/models/isolf_new.html}}. \citet{dartmouth} calculations employ a metal distribution with [$\alpha$/Fe]=$-$0.20 as in our models, although the reference solar mixture \citep[the][solar metal composition is used by these authors]{gres} is not the same, and there are differences also in some physics inputs \citep[see][for details]{dartmouth}.

Apart from the lower~MS, the two sets of isochrones are in close 
agreement. Differences in $T_{\rm eff}$ at fixed luminosity are 
at most by a few tens of K, and the TO luminosities are always practically the same, apart from the 1 and 5~Gyr, [Fe/H]=$-$1.9 cases, 
with the \citet{dartmouth} isochrones showing a slightly brighter TO than our calculations, corresponding to an age 
difference by about 100~Myr.

The disagreement between the two sets of models along the fainter portion of the MS is due to the use of different approaches for the outer boundary conditions,  an input that significantly affects the $T_{\rm eff}$ scale of very low-mass stars, as well as to differences in the adopted chemical mixtures (see below for more details).

\section{The role of the bolometric corrections}
\label{bolom}

As shown by \citet{bcaen} and confirmed by our more recent calculations \citep{bastiiacae}, $\alpha$-enhanced isochrones with ages above $\sim$1~Gyr can be well mimicked by solar-scaled ones with the same total metallicity [M/H] 
not only in the HRD, but also in $VI$ and infrared colour magnitude diagrams (CMDs), 
whilst the agreement is less satisfactory in optical CMDs and worsens at shorter wavelengths. 
The good match in the HRD confirms the results of the pioneering study by  \citet{saen} who at the time did not have available bolometric corrections (BCs) calculated with a consistent $\alpha$-enhanced metal distribution.
Indeed, the poorer agreement in CMDs at optical and shorter wavelengths is due to differences in the BCs induced by the effect of the change in the metal abundance distribution on the stellar spectra.

\begin{figure}
	\includegraphics[width=\columnwidth]{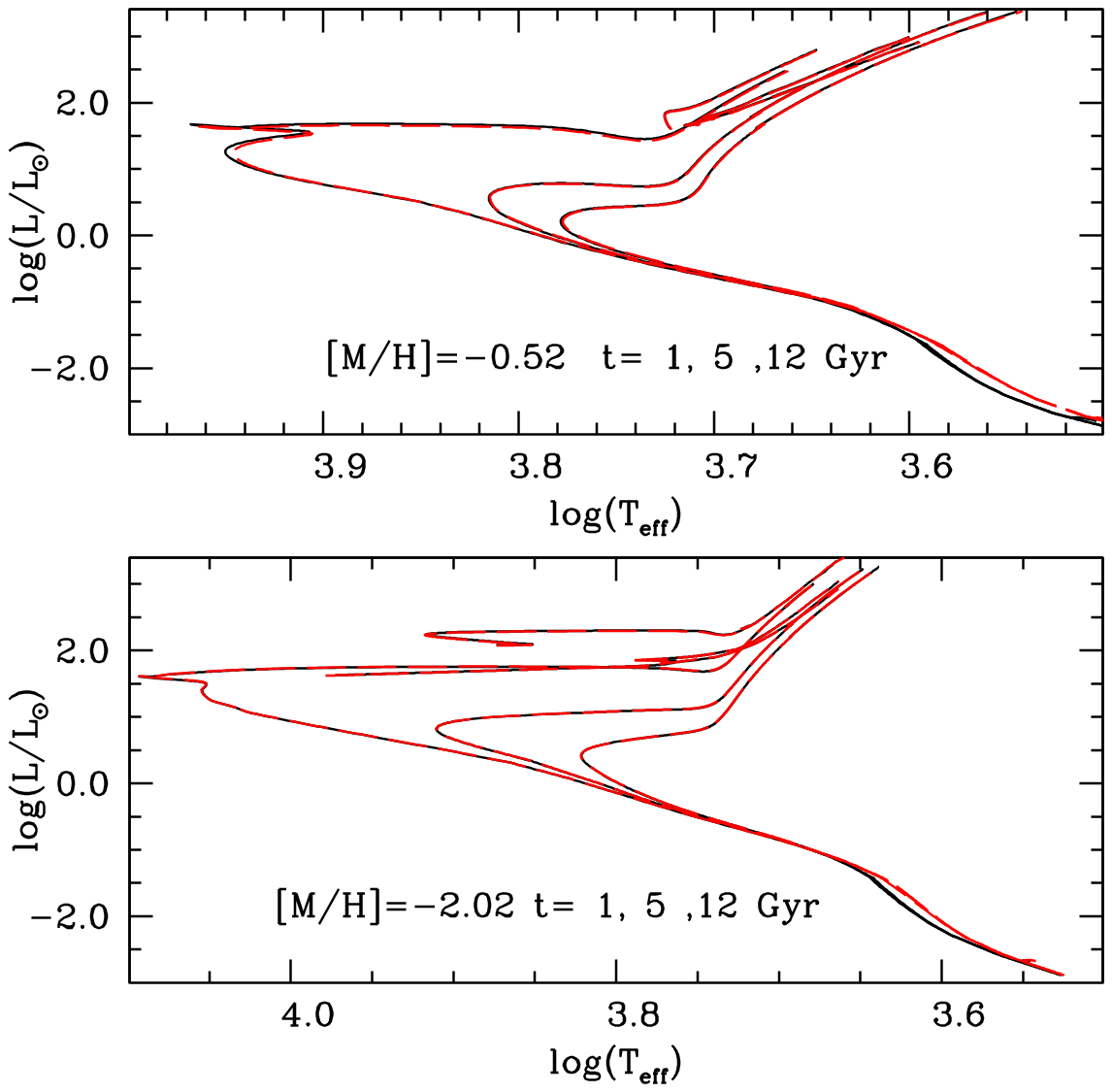}
        \caption{Comparison in the HRD between our $\alpha$-depleted (black solid lines) and solar-scaled 
        \citep{bastiiacss} isochrones (red dashed lines) with the labelled total metallicities  [M/H] and ages.}
    \label{fig_b}
\end{figure}

We expect that similar results should hold also when comparing $\alpha$-depleted with solar-scaled isochrones at the same [M/H], and indeed this is the case, as shown in Fig.~\ref{fig_b}.

\begin{figure}
	\includegraphics[width=\columnwidth]{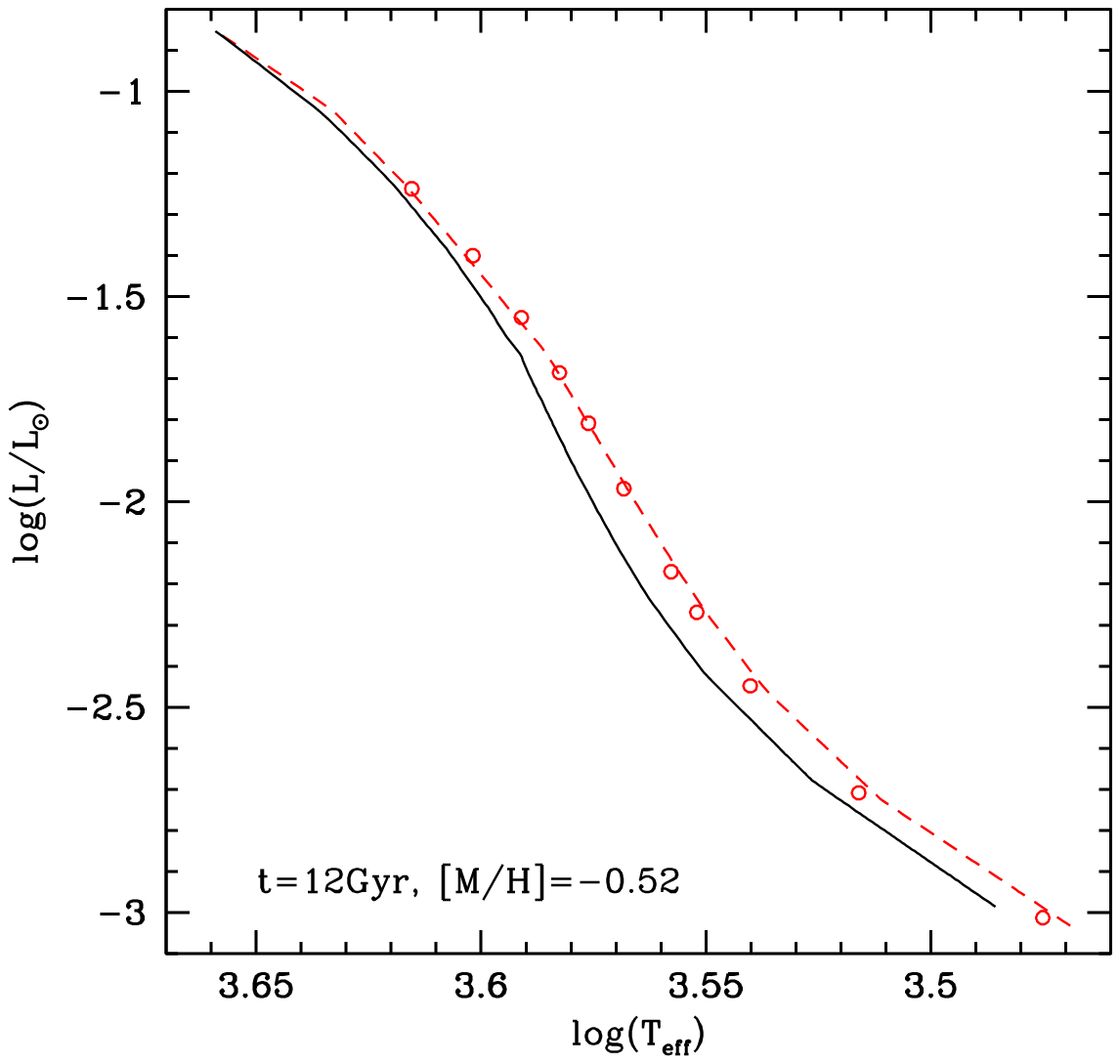}
        \caption{Comparison in the HRD between $\alpha$-depleted (black solid lines) and solar-scaled 
        \citep{bastiiacss} isochrones (red dashed lines) for the labelled values of the total metallicity and age. The open dots denote solar-scaled models with mass between $0.1~M_\odot$ and $0.5~M_\odot$ for the same values of [M/H] and age, but computed employing for the outer boundary conditions the same $T(\tau)$ relationship adopted for the $\alpha-$depleted calculations.}
    \label{fig:vlm}
\end{figure}

Here we compare our $\alpha$-depleted 1, 5, and 12~Gyr isochrones with the solar-scaled ones from \citet{bastiiacss} at two different common values of [M/H]. There is a perfect match between the two pairs of isochrones along MS, TO, and the core He-burning phase after the He-flash at the tip of the RGB. Differences appear only on the lower MS for log$(L/L_{\odot})$ below $\sim -$1.0 corresponding to masses below $\sim$ 0.5$M_{\odot}$. Similar differences for the very low-mass regime were found in Paper~II when comparing solar-scaled (from Paper~I) with $\alpha-$enhanced models. To verify the role played by the outer boundary conditions in causing 
these differences, we computed solar-scaled stellar models in this mass regime for [M/H]=$-$0.52, by adopting the same $T(\tau)$ relationship used for our $\alpha$-depleted computations. Figure~\ref{fig:vlm} shows the resulting 12~Gyr isochrone compared to our new $alpha$-depleted calculations, and a solar-scaled one 
from Paper~I models.
When comparing the two solar-scaled isochrones   
it is easy to notice that the use of a 
$T(\tau)$ relationship instead of the more appropriate boundary conditions provided by model atmospheres  employed in Paper~I calculations, makes 
the very low-mass models hotter \citep[see also][and references therein]{baraffe, cassisivlm}. However, the impact is not able to explain the full difference between our solar-scaled calculations and the $\alpha-$depleted ones at the same [M/H] (the same is 
true for the differences with the $\alpha-$enhanced models shown in Paper~II). Indeed the main source of 
these $T_{\rm eff}$ differences is not so much the way the boundary conditions are computed, rather the metal 
mixture used in their calculations.
As already discussed by \cite{don:22} and \cite{don:23}, 
the boundary conditions are 
much more sensitive to the adopted specific metal abundance distribution 
than for more massive models. In particular, \cite{don:23} has demonstrated the crucial role played by oxygen -- whose abundance changes substantially when moving from the solar-scaled mixture to the $\alpha-$depleted/enhanced ones -- in determining the atmospheric thermal stratification of very low-mass stars. The differences in the boundary conditions 
for very low-mass models at a given total metallicity are the primary 
reason for the $T_{\rm eff}$ differences between 
solar-scaled and $\alpha$-depleted isochrones shown in Fig.~\ref{fig:vlm}.

\begin{figure}
\includegraphics[width=\columnwidth]{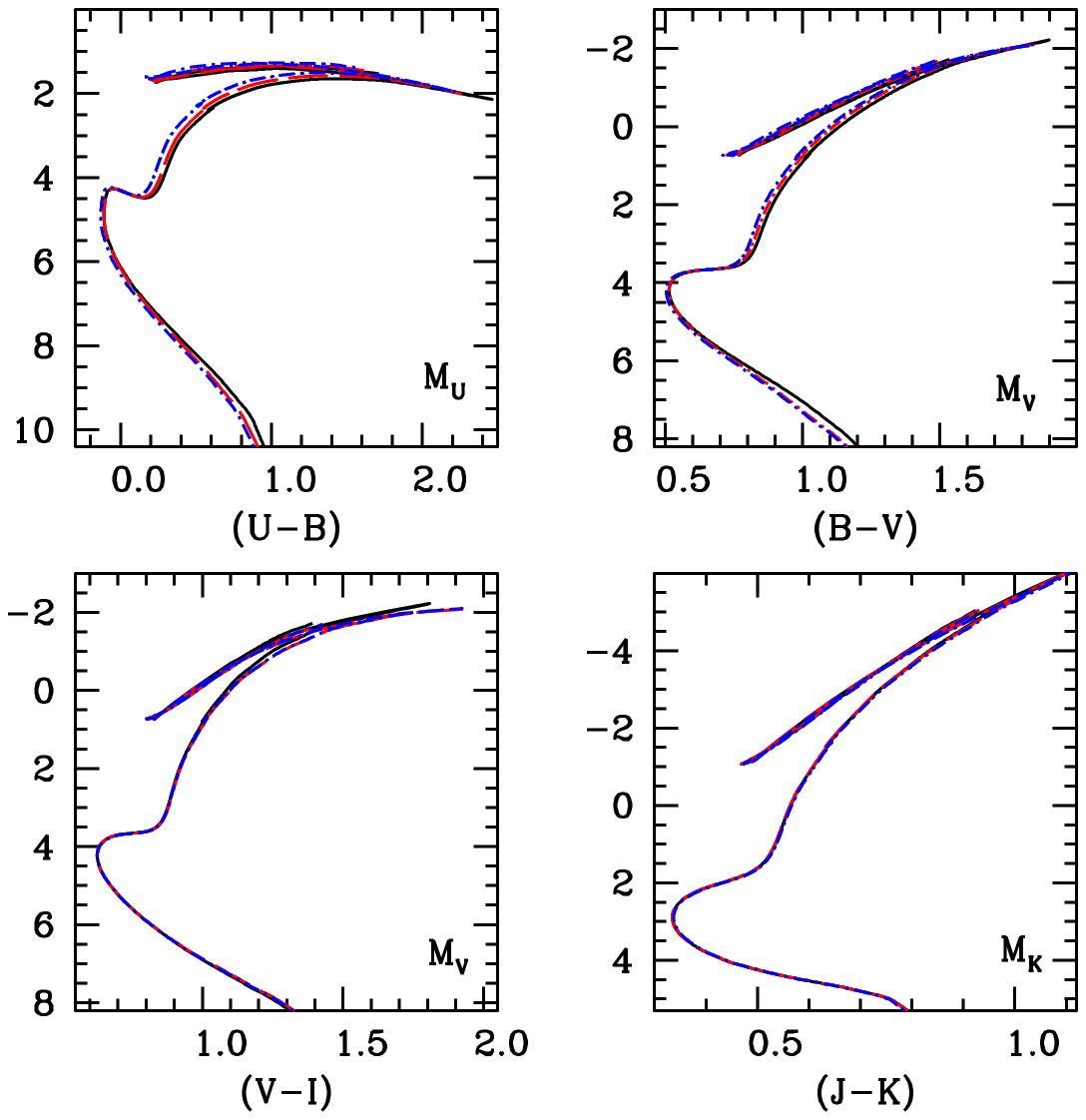}
\caption{Comparison in various CMDs of a self-consistent 12~Gyr, [Fe/H]=$-$0.7 $\alpha$-depleted isochrone (black solid lines), with isochrones for the same age and chemical composition but calculated using solar-scaled bolometric corrections with either the same [Fe/H] (red dashed lines) or with the same [M/H] (blue dash-dotted lines) of the self-consistent isochrone, respectively. We didn't display the portion of the isochrones populated by masses below $\sim$0.5$M_{\odot}$(see text for details).}
\label{fig_d}
\end{figure}

Figure~\ref{fig_d} highlights the effect of the $\alpha$-depleted 
heavy element distribution of the model BCs, the same way it was shown 
by \citet{bcaen} for the $\alpha$-enhanced case. We display here 
12~Gyr, [Fe/H]=$-$0.7 $\alpha$-depleted isochrones in different 
CMDs, calculated using the appropriate BCs for a consistent $\alpha$-depleted metal mixture (we denote them as self-consistent isochrones), 
BCs calculated for a solar-scaled metal mixture and the same [Fe/H] 
of the isochrone (denoted as [Fe/H] isochrones) and BCs computed 
for a solar-scaled metal mixture and the same [M/H] 
of the isochrone (denoted as [M/H] isochrones), respectively.
We do not show the lower MS, for which we do not have models computed using boundary conditions from $\alpha$-depleted model atmospheres.

As in \citet{bcaen} we find that the perfect agreement between the three 
isochrones in the infrared CMD progressively worsens, especially at 
low $T_{\rm eff}$, when moving to CMDs at increasingly shorter wavelenghts.
Moreover, these differences quantitatively decrease with decreasing [Fe/H] of the self-consistent $\alpha$-depleted isochrone

Just for reference, in the $UB$ CMD the RGB at $M_U$=3 for the [M/H] isochrone is 0.08~mag bluer than the self-consistent one, while  
the [Fe/H] solar-scaled isochrone is 0.03~mag bluer.
The absolute $U$ magnitude and $(U-B)$ colour of the TO are the same for the self-consistent and the [Fe/H] isochrone, while $M_U$ is $\sim$0.03~mag fainter and $(U-B)$ is 0.02~mag bluer for the [M/H] isochrone.
On the MS at $M_U$=9 the [M/H] isochrone is 0.07~mag bluer than the self-consistent one, while  
the [Fe/H] solar-scaled isochrone is 0.04~mag bluer. These differences 
are substantially larger when considering magnitudes and colours in the UV 
wavelength range.

Moving to the $BV$ CMD the differences decrease. On the RGB at $M_V$=1 the [M/H] isochrone is now 0.04~mag bluer than the self-consistent one, while 
the [Fe/H] isochrone is 0.02~mag bluer. 
The TO colours and magnitudes are basically identical among the three 
isochrones, while on the MS at $M_V$=7.5 the [M/H] isochrone is 0.06~mag bluer than the self-consistent one and 
the [Fe/H] solar-scaled isochrone is 0.05~mag bluer.
In the $VI$ CMD differences appear only when approaching the tip of the RGB,  
which is redder in both [Fe/H] and [M/H] isochrones by about 0.12~mag. 

On average, the [Fe/H] isochrone is a better approximation of the self-consistent one, in agreement with the results by \citet{bcaen} for the $\alpha$-enhanced regime.
Not surprisingly, the sign of the colour differences is the opposite of the $\alpha$-enhanced comparison, whereby the [Fe/H] and [M/H] isochrones are  redder along the RGB and the MS.
As discussed briefly in \citet{bcaen}, the different abundances of 
mainly O, Si, Mg, and Ca between $\alpha$-depleted and solar-scaled compositions that affect both the line opacity and the continuum opacity 
in the atmospheres, causing a redistribution of the flux among the various wavelengths, and a change of the values of the BCs at fixed $T_{\rm eff}$ and gravity.

\section{Summary}
\label{concl}

We have presented our new BaSTI $\alpha$-depleted ([$\alpha$/Fe]=$-$0.20) 
stellar models 
and isochrones, suitable to investigate the evolutionary properties 
of the $\alpha$-depleted populations discovered in the Milky Way and Local Group 
dwarfs.
The models have been 
calculated with the same physics inputs and reference solar
metal mixture employed for our updated solar-scaled and $\alpha$-enhanced ([$\alpha$/Fe]=0.4) calculations published in Paper~I and II, and cover the same mass and [Fe/H] ranges.

We have compared the HRDs of selected $\alpha$-depleted isochrones with the \citet{dartmouth} [$\alpha$/Fe]=$-$0.20 counterparts, from the MS to the tip of the RGB, finding an overall good agreement despite differences in
the reference solar mixture and some physics inputs. We have also discussed 
the importance of employing bolometric corrections calculated for the 
appropriate $\alpha$-depleted heavy element mixture when comparing isochrones 
(and evolutionary tracks) to data in optical CMDs and at shorter wavelengths.
 
Like for the solar-scaled and $\alpha$-enhanced libraries, these 
new $\alpha$-depleted evolutionary tracks and isochrones are available 
in several photometric systems at \url{http://basti-iac.oa-abruzzo.inaf.it}. In the near future, we will make available on the BaSTI website an online tool to interpolate among our sets of calculations, to obtain tracks ad isochrones in all available photometric filters for any    
value of [$\alpha$/Fe] between $-$0.2 and +0.4.

\section*{Acknowledgements}
We warmly thank our anonymous referee for 
pointing out the origin of the differences 
between our sets of very low-mass models computed  
with different heavy element mixtures. SC and AP acknowledge support from PRIN-MIUR2022 (PI: S. Cassisi), from INAF Theory grant "Lasting" (Responsabile: S. Cassisi),
INFN (Iniziativa specifica TAsP), and from PLATO ASI-INAF agreement
n.2015-019-R.1-2018. 
MS acknowledges support from STFC Consolidated Grant ST/V00087X/1.

\section*{Data Availability}

The $\alpha$-depleted evolutionary tracks and isochrones in several photometric systems are available at the new 
official BaSTI website \url{http://basti-iac.oa-abruzzo.inaf.it}.



\bibliographystyle{mnras}
\bibliography{am02paper} 

\begin{table*}
\caption{Available photometric systems. For each system, information about the source from the response curves and reference zero-points is provided.}
\begin{threeparttable}
\label{tab:filters}
\begin{tabular} {llll}
\toprule
Photometric system & Calibration & Passbands & Zero-points\\
 \midrule
2MASS & Vegamag & \citet{Cohen03}  &\citet{Cohen03}\\
CFHT (MegaCam) & ABmag& CFHT Documentation\tnote{1}& 0\\

Euclid (VIS + NISP) & ABmag & Euclid mission database\tnote{2} & 0\\
Gaia DR1& Vegamag & \citet{Jordi10}\tnote{3}  &\citet{Jordi10}\\
Gaia DR2 & Vegamag &\citet{Maiz-Apellaniz18}\tnote{4}&\citet{Maiz-Apellaniz18}\\
Gaia DR3 & Vegamag & \citet{riello} & \citet{riello}\\
GALEX & ABmag & NASA\tnote{5} & 0\\
Hipparcos + Thyco&ABmag&\citet{Bessell12}&\citet{Bessell12}\\
HST (WFPC2) & Vegamag & SYNPHOT &SYNPHOT\\
HST ( WFC3) & Vegamag & SYNPHOT &SYNPHOT\\
HST (ACS) & Vegamag & SYNPHOT  &SYNPHOT\\
Johnson, Bessel \& Brett & Vegamag & \citet{Bessel88,Bessel90}  &\citet{Bessel98}\\
J-PLUS& ABmag& J-PLUS collaboration \tnote{6}& 0 \\
JWST (NIRCam - Post Launch) & Vegamag & JWST User Documentation \tnote{7}&SYNPHOT\\
JWST (NIRCam - Sirius) \tnote{7a} & Vegamag & JWST User Documentation \tnote{7}&SYNPHOT\\
JWST (NIRISS - Post Launch) & Vegamag & JWST User Documentation \tnote{7}&SYNPHOT\\
Kepler& ABmag&Kepler collaboration\tnote{8}&0\\
Nancy Grace Roman Telescope (WFI) & Vegamag & Roman Documentation \tnote{9}& SYNPHOT\\
PanSTARSS1 & ABmag & \citet{Tonry12} & 0\\
SAGE & ABmag &  SAGE collaboration & 0\\
SDSS & ABmag &  \citet{Doi10} &\citet{Dotter08}\\
Skymapper & ABmag & \citet{Bessell11}  & 0\\
Spitzer (IRAC) & Vegamag & NASA\tnote{10} & \citet{Groenewegen06}\\
Str\"{o}mgren & Vegamag & \citet{Maiz-Apellaniz06}   &\citet{Maiz-Apellaniz06}\\
Subaru (HSC) & ABmag & HSC collaboration\tnote{11} & 0\\
SWIFT (UVOT) & Vegamag & HEASARC\tnote{12} & \citet{Poole08}\\
TESS&ABmag&TESS collaboration\tnote{13}& 0\\
UVIT (FUV + NUV + VIS)&ABmag&UVIT collaboration \tnote{14}&\citet{Tandon17}\\ 
Vera C. Rubin Observatory & ABmag & LSST collaboration \tnote{15}& 0\\
Victor M. Blanco Telescope (DECam)  & ABmag & CTIO \tnote{16} & 0\\
VISTA (VIRCAM) & Vegamag &  ESO\tnote{17} & \citet{Rubele12}\\
WISE & Vegamag & WISE collaboration\tnote{18} & \citet{Wright10}\\
\bottomrule

\end{tabular}

\begin{tablenotes}
\item[1]{\url{https://www.cfht.hawaii.edu/Instruments/Filters/megaprimenew.html}}
\item[2]{\url{http://euclid.esac.esa.int/epdb/db/SPV02/SPV02/EUC_MDB_MISSIONCONFIGURATION-SPV02_2018-06-16T140000.00Z_01.01.xml.html}}
\item[3]{The nominal G passband curve has been corrected following the post-DR1 correction provided by \citet{Maiz-Apellaniz17}}
\item[4]{Two different $G_{BP}$ passbands are provided for sources brighter and fainter than G=10.87, respectively.}
\item[5]{\url{https://asd.gsfc.nasa.gov/archive/galex/Documents/PostLaunchResponseCurveData.html}}
\item[6]{\url{http://www.j-plus.es/survey/instrumentation}}
\item[7]{\url{https://jwst-docs.stsci.edu/}}
\item[7a]{This photometric system adopts the "post launch passbands but Vega zeropoints obtained by using Sirius as the color reference (see JWST User Documentation for more details).}
\item[8]{\url{https://nexsci.caltech.edu/workshop/2012/keplergo/CalibrationResponse.shtml}}
\item[9]{\url{https://roman.gsfc.nasa.gov/science/WFI_technical.html}}
\item[10]{\url{https://irsa.ipac.caltech.edu/data/SPITZER/docs/irac/calibrationfiles/spectralresponse/}}
\item[11]{\url{https://hsc-release.mtk.nao.ac.jp/doc/index.php/survey/}}
\item[12]{\url{https://heasarc.gsfc.nasa.gov/docs/heasarc/caldb/data/swift/uvota/index.html}}
\item[13]{\url{https://heasarc.gsfc.nasa.gov/docs/tess/data/tess-response-function-v1.0.csv}}
\item[14]{\url{https://uvit.iiap.res.in/Instrument/Filters}}
\item[15]{\url{https://github.com/lsst/throughputs/tree/master/baseline}}
\item[16]{\url{http://www.ctio.noirlab.edu/noao/content/decam-filter-information}}
\item[17]{\url{http://www.eso.org/sci/facilities/paranal/instruments/vircam/inst/}}
\item[18]{\url{https://wise2.ipac.caltech.edu/docs/release/prelim/expsup/sec4_3g.html#WISEZMA}}

\end{tablenotes}
\end{threeparttable}
\end{table*}
\clearpage








\bsp	
\label{lastpage}
\end{document}